\documentclass{aastex631}
\usepackage{graphicx}
\usepackage{amsmath}
\usepackage{natbib}
\graphicspath{{figures/}}
\usepackage{textcomp}
\begin{document}

\shorttitle{3I/ATLAS Post-Perihelion Activity}
\shortauthors{Scarmato \& Loeb}

\title{Periodic Wobble of the Post-Perihelion Jet Structure Around 3I/ATLAS}

\author{Toni Scarmato}
\affiliation{Toni Scarmato's Astronomical Observatory, San Costantino di Briatico, Calabria, 89817, Italy}

\author{Abraham Loeb}
\affiliation{Astronomy Department, Harvard University, 60 Garden Street, Cambridge MA 02138, USA}

\begin{abstract}
We analyze data on the post-perihelion morphology, including jet position angles (PAs) and coma-dominated photometry of the interstellar object 3I/ATLAS.
From \textit{Hubble Space Telescope} (HST) images processed with a Larson--Sekanina rotational-gradient filter, we measure the PAs of three main persistent jet-like features between November 30 and December 27, 2025 and fit a weighted Fourier model in a period scan.
The dominant jet (Jet 2) PA wobble yields $P_{\rm jet}=7.20\pm0.05$~h. The other two jets (Jet 1 and Jet 3) oscillate with periods of 2.9 h and 4.3 h, respectively. Significant is the result of their sum, which is exactly equal to the period of the main jet (Jet 2). 
Moreover, an independent $G_r$ ($\approx R$) time-series photometry data set, using two different apertures (4 pixels and 30 pixels) from MPC station L92, analyzed with nightly offsets and 30-minute binning, gives $P_{\rm phot}=7.136\pm0.001$~h ($1\sigma$), with a semi-amplitude $A\simeq0.311$~mag and scatter $\sigma_{\rm jit}\simeq0.089$~mag.
The close agreement between the periods supports a characteristic post-perihelion period of $\sim7.1$~h.
We interpret this period as an attitude precession/nutation (non-principal-axis rotation) traced by jet orientation and coma flux redistribution. The main jet structure precesses about the rotation axis with a characteristic angular excursion of order $\sim20^\circ$, and the rotation axis is aligned with the sunward direction to within $\sim20^\circ$. The agreement between independent morphological and photometric tracers supports a stable post-perihelion activity period near 7.1 h.
\end{abstract}


\section{Introduction}
The rotation of comets nuclei (de La Fuente Marcos et al. 2005) affects the temporal modulation of their outgassing, the formal persistence of jets, and the evolution of coma morphology (Hacque and Lopez 2025); (Eubanks et al. 2025). Rotation periods are commonly inferred either from time-resolved photometry (reflecting changes in projected cross section and rotating activity pattern) or from morphological tracers such as repeating jet or spiral features whose apparent geometry is modulated by nucleus rotation and viewing geometry (e.g. Samarasinsha et al. 2024; Farnham and Schleicher 2005). Both diagnostics can be affected by aliasing (sigle-vs.-double peak modulation), evolving activity, and non-Gaussian systematics arising from coma contamination, variable seeing, and calibration residuals. A robust period inference uses photometry and morphology as complementary constraints (Serra-Ricart et al. 2025). 
Jets are among the most direct probes of cometary nucleus activity.
Time-variable jet position angles (PAs) and coma brightness modulations can constrain rotation, precession, and the distribution of active areas \citep[e.g.,][]{LarsonSekanina1984,Jewitt1997,Samarasinha2004,FarnhamSchleicher2005}.
Coma-dominated photometry does not necessarily trace the spin of a bare nucleus: instead, it often reflects a time-dependent mapping from jet geometry and dust production into the chosen aperture.
Non-principal-axis (NPA) rotation and precession can produce quasi-periodic PA ``wobbles'' and non-sinusoidal waveforms, particularly when the coma evolves into fan-like structures \citep[e.g.,][]{SamarasinhaAhEarn1991,Belton2010}.

For unevenly sampled time series, period searches commonly rely on the Lomb-Scargle formalism and its generalizations \citep{Lomb1976,Scargle1982,Zechmeister2009}. 
The Lomb-Scargle formalism (Lomb 1976) and its generalizations provide standard approaches for identifying periodic signals and evaluating their significance under irregular sampling (Lomb1976; Scargle 1982; Zechmeister and Kürster 2009).
When the observable is non-sinusoidal, truncated Fourier-series representations provide a compact model solvable by linear least squares at fixed trial period \citep[e.g.,][]{Press2007}.

In a previous paper \citep{ScarmatoLoeb2026} we derived: (i) a jet-PA period scan from the Hubble Space Telescope (HST) image morphology and (ii) an independent ground-based photometric period search to constrain the activity and period of 3I/ATLAS.
That analysis identified a jet-PA wobble period $P_{\rm jet}=7.20\pm0.05$~h from HST epochs spanning November 30--December 27, 2025 and a photometric period $P_{\rm phot}=7.136\pm0.001$~h from 30-minute binned $G_r$-band photometry obtained on four nights.
The periods are not identical but their similarity is consistent with non-Gaussian systematics, aliasing, and the different transfer functions between nucleus attitude, jet morphology, and coma-integrated flux.
Our goal here is to interpret the inferred $\sim7.1$~h period explicitly in terms of an attitude precession/nutation signature, and to connect it to multi-jet geometry and aperture-dependent photometric responses as the inner coma transitions between collimated and fan-like morphologies.

Here we apply harmonic modeling to a phase-folded jet-PA time series for secondary jets of 3I/ATLAS between November 30, 2025 and December 27, 2025, and compare with time series observations between December 9, 2025 and December 22, 2025, using two apertures,  to determine the period of the wobbling post-perihelion jet structure around 3I/ATLAS (Santana-Ros et al. 2025).  
\
\section{Observations and Data}
\subsection{\textbf{Morphology and jet PA measurements}}
Within the coma, we find evidence for three jets, separated equally from each other, by $\approx$ 120° in the sky, in addition to the anti-tail outflow which extends toward the Sun out to a much longer distance. PAs are measured east of north in the plane of the sky.
Jet~2 is approximately anti-sunward and close to the projected spin-axis direction inferred from the global morphology.
Figure~\ref{fig:schematic} summarizes the geometry and the mean jets PAs with their measured oscillation half-amplitudes. 
The jet position angle (PA) time series used to determine the rotation period was derived from \textit{Hubble Space Telescope} (\textit{HST}) images by enhancing low-contrast coma structures through a Larson--Sekanina (LS) filter. LS processing is widely used in cometary morphology studies to suppress the quasi-radial coma component and amplify azimuthal intensity gradients associated with jets, fans, and other anisotropic outflow features.

For each \textit{HST} image, the calibrated frames were registered to a common centroid and (when multiple exposures were available) combined to improve the signal-to-noise ratio while preserving small-scale morphology. The LS operator was applied as a rotational differential filter of the form
\begin{equation}
I_{\rm LS}(r,\theta) = I(r,\theta) - I(r,\theta+\Delta\theta),
\end{equation}
with the rotation increment $\theta$ = 31 ° and $\delta$r = 0.1 chosen to maximize the contrast of the primary jet while avoiding the introduction of spurious multiple-lobe artifacts. In practice, we experimented with a narrow range of $\Delta\theta$ values and adopted the setting that produced the most stable morphology between frames within the same epoch. The resulting enhanced images were inspected to identify the dominant jet ridge line.
The jet PA was measured in the conventional sense (east of north), by determining the orientation of the enhanced jet axis relative to the photo-center. Measurements were obtained by tracing the jet ridge line over a fixed radial range from the nucleus (chosen to minimize saturation/PSF residuals near the core and to avoid low-S/N regions at large radii) and fitting a straight line to the ridge in the sky plane. The best-fitting orientation was converted to PA, and its uncertainty was estimated by propagating the dispersion among repeated measurements (e.g., varying the radial range and LS parameters within a conservative interval) and by accounting for the finite angular resolution of the enhanced jet feature. The final PA uncertainties used in the period fitting are reported in Table~1.

\subsection{\textbf{Jet PA Measurement Procedure}}
\textbf{To ensure reproducibility of the jet position angle (PA) measurements, we explicitly describe the method used. After applying the Larson–Sekanina filter, the main jet structure is identified in the enhanced image. The jet ridge line is traced within a fixed radial range selected to avoid PSF-dominated regions near the nucleus and low signal-to-noise regions at large distances.
The PA measurements were performed directly on the Larson–Sekanina enhanced images by overlaying a calibrated angular grid centered on the photo-center of the comet. The initial jet direction was identified from the first five brightest pixels emerging from the nucleus. The measured direction was then verified by following the jet ridge line over approximately 30 pixels from the nucleus. Independent measurements performed on different image stretches and enhancement parameters produced consistent results. Considering the WFC3/UVIS pixel scale (0.0395 arcsec/pixel), the typical HST PSF of  about 0.07 arcsec (about 2 pixels), and the repeatability of the measurements, a conservative uncertainty of +/-3° was adopted for all PA determinations.}
\textbf{The final uncertainty is conservatively estimated as +/-3°, accounting for the finite width of the jet ridge, the WFC3/UVIS PSF, the repeatability of the manual angular alignment, and the stability of the measured direction over the  about 30 pixel radial extent.}

\begin{deluxetable}{lccccccc}
\tablecaption{Main (\textbf{Jet 2}) and secondary (\textbf{Jet 1 and Jet 3})  position angle measurements.\label{tab:pa}}
\tablehead{\colhead{\textbf{UT date}} & \colhead{\textbf{UTC Time}} & \colhead{\textbf{PA (Jet 2})} & \colhead{\textbf{PA (Jet 1})} & \colhead{\textbf{PA (Jet 3)}}& \colhead{\textbf{err ($3\sigma$})} & \colhead{\textbf{$\Delta$t (h)}}  }
\startdata
2025 11 30.80903 & 2025-11-30 20:36:57 & 290 & 65 & 174 & +/-3° & 0.000\\
2025 11 30.86389 & 2025-11-30 20:44:00 & 288 & 62 & 169 & +/-3° & 1.317\\
2025 11 30.86875 & 2025-11-30 20:51:00 & 288 & 58 & 164 & +/-3° & 1.433\\
2025 11 30.87431 & 2025-11-30 20:59:00 & 285 & 53 & 162 & +/-3° & 1.567\\
2025 11 30.87911 & 2025-11-30 21:06:39 & 283 & 49 & 161 & +/-3° & 1.682\\
2025 11 30.88403 & 2025-11-30 21:13:56 & 283 & 45 & 161 & +/-3° & 1.800\\
2025 12 04.65068 & 2025-12-04 15:37:22 & 259 & 49 & 161 & +/-3° & 92.200\\
2025 12 04.65554 & 2025-12-04 15:44:49 & 262 & 50 & 158 & +/-3° & 92.316\\
2025 12 04.66110 & 2025-12-04 15:52:50 & 265 & 52 & 157 & +/-3° & 92.450\\
2025 12 04.66666 & 2025-12-04 16:00:17 & 268 & 54 & 155 & +/-3° & 92.583\\
2025 12 12.88889 & 2025-12-12 21:20:32 & 281 & 53 & 165 & +/-3° & 289.917\\
2025 12 12.89306 & 2025-12-12 21:26:28 & 278 & 52 & 164 & +/-3° & 290.017\\
2025 12 12.89721 & 2025-12-12 21:32:24 & 277 & 52 & 163 & +/-3° & 290.117\\
2025 12 12.90139 & 2025-12-12 21:38:54 & 275 & 51 & 163 & +/-3° & 290.217\\
2025 12 12.90556 & 2025-12-12 21:44:50 & 273 & 50 & 162 & +/-3° & 290.317\\
2025 12 12.90971 & 2025-12-12 21:50:46 & 270 & 50 & 162 & +/-3° & 290.417\\
2025 12 27.66528 & 2025-12-27 15:58:33 & 263 & 66 & 186 & +/-3° & 644.550\\
2025 12 27.66943 & 2025-12-27 16:04:29 & 260 & 65 & 188 & +/-3° & 644.650\\
2025 12 27.67431 & 2025-12-27 16:11:00 & 258 & 62 & 189 & +/-3° & 644.767\\
2025 12 27.67993 & 2025-12-27 16:16:56 & 256 & 60 & 192 & +/-3° & 644.902\\
2025 12 27.68193 & 2025-12-27 16:22:52 & 255 & 57 & 195 & +/-3° & 644.950\\
\enddata
\tablecomments{Dates are expressed as YYYY~MM~DD.ddddd. The uncertainty is conservatively taken as 3°.}
\end{deluxetable}

\begin{deluxetable*}{lllllllll}
\tablecaption{HST WFC3/UVIS observations of 3I/ATLAS. All observations were obtained with the HST WFC3/UVIS channel using the F350LP filter. The data include observations from multiple programs (PI: Hui and Jewitt), covering the post-perihelion evolution of 3I/ATLAS.}
\tablehead{
\colhead{Dataset} & \colhead{Target} & \colhead{Instrument} & \colhead{Aperture} & \colhead{Filter} & \colhead{Exp. Time (s)} & \colhead{Start Time (UTC)} & \colhead{RA (J2000)} & \colhead{Dec (J2000)}
}
\startdata
IFLE01FBQ & 3I-ATLAS-OCT10 & WFC3 & UVIS2-2K2C-SUB & F350LP & 260 & 2025-11-30 20:36:57 & 180.5647976 & 0.4177606 \\
IFLE01FCQ & 3I-ATLAS-OCT10 & WFC3 & UVIS2-2K2C-SUB & F350LP & 260 & 2025-11-30 20:44:14 & 180.5599871 & 0.4194685 \\
IFLE01FDQ & 3I-ATLAS-OCT10 & WFC3 & UVIS2-2K2C-SUB & F350LP & 260 & 2025-11-30 20:51:31 & 180.5555045 & 0.4210601 \\
IFLE01FFQ & 3I-ATLAS-OCT10 & WFC3 & UVIS2-2K2C-SUB & F350LP & 260 & 2025-11-30 20:59:22 & 180.5506012 & 0.4228085 \\
IFLE01FGQ & 3I-ATLAS-OCT10 & WFC3 & UVIS2-2K2C-SUB & F350LP & 260 & 2025-11-30 21:06:39 & 180.5461176 & 0.4244004 \\
IFLE01FHQ & 3I-ATLAS-OCT10 & WFC3 & UVIS2-2K2C-SUB & F350LP & 260 & 2025-11-30 21:13:56 & 180.5416336 & 0.4259925 \\
IFKP04DLQ & 3I-ATLAS-OCT29 & WFC3 & UVIS2-2K2C-SUB & F350LP & 270 & 2025-12-04 15:37:22 & 177.0945289 & 1.6542930 \\
IFKP04DMQ & 3I-ATLAS-OCT29 & WFC3 & UVIS2-2K2C-SUB & F350LP & 270 & 2025-12-04 15:44:49 & 177.0892655 & 1.6561667 \\
IFKP04DNQ & 3I-ATLAS-OCT29 & WFC3 & UVIS2-2K2C-SUB & F350LP & 270 & 2025-12-04 15:52:50 & 177.0839073 & 1.6580793 \\
IFKP04DOQ & 3I-ATLAS-OCT29 & WFC3 & UVIS2-2K2C-SUB & F350LP & 270 & 2025-12-04 16:00:17 & 177.0789945 & 1.6598282 \\
IFLE22F7Q & 3I-ATLAS-DEC1 & WFC3 & UVIS2-2K2C-SUB & F350LP & 170 & 2025-12-12 21:20:32 & 168.6982788 & 4.6255238 \\
IFLE22F8Q & 3I-ATLAS-DEC1 & WFC3 & UVIS2-2K2C-SUB & F350LP & 170 & 2025-12-12 21:26:28 & 168.6934067 & 4.6272314 \\
IFLE22F9Q & 3I-ATLAS-DEC1 & WFC3 & UVIS2-2K2C-SUB & F350LP & 170 & 2025-12-12 21:32:24 & 168.6889361 & 4.6287982 \\
IFLE22FAQ & 3I-ATLAS-DEC1 & WFC3 & UVIS2-2K2C-SUB & F350LP & 170 & 2025-12-12 21:38:54 & 168.6839654 & 4.6305444 \\
IFLE22FBQ & 3I-ATLAS-DEC1 & WFC3 & UVIS2-2K2C-SUB & F350LP & 170 & 2025-12-12 21:44:50 & 168.6794943 & 4.6321114 \\
IFLE22FCQ & 3I-ATLAS-DEC1 & WFC3 & UVIS2-2K2C-SUB & F350LP & 170 & 2025-12-12 21:50:46 & 168.6750230 & 4.6336784 \\
IFLE03GEQ & 3I-ATLAS-DEC1 & WFC3 & UVIS2-2K2C-SUB & F350LP & 170 & 2025-12-27 15:58:33 & 151.4682138 & 10.3650160 \\
IFLE03GFQ & 3I-ATLAS-DEC1 & WFC3 & UVIS2-2K2C-SUB & F350LP & 170 & 2025-12-27 16:04:29 & 151.4627892 & 10.3666960 \\
IFLE03GGQ & 3I-ATLAS-DEC1 & WFC3 & UVIS2-2K2C-SUB & F350LP & 170 & 2025-12-27 16:11:00 & 151.4572456 & 10.3684093 \\
IFLE03GHQ & 3I-ATLAS-DEC1 & WFC3 & UVIS2-2K2C-SUB & F350LP & 170 & 2025-12-27 16:16:56 & 151.4522685 & 10.3699506 \\
IFLE03GIQ & 3I-ATLAS-DEC1 & WFC3 & UVIS2-2K2C-SUB & F350LP & 170 & 2025-12-27 16:22:52 & 151.4472914 & 10.3714919 \\
\enddata
\end{deluxetable*}

\textbf{This approach yields a homogeneous PA time series from space-based imaging that is minimally affected by seeing variations and that directly traces the rotational modulation of the active source region driving the observed jet morphology.}
\begin{figure*}
\centering
\includegraphics[width=0.95\textwidth]{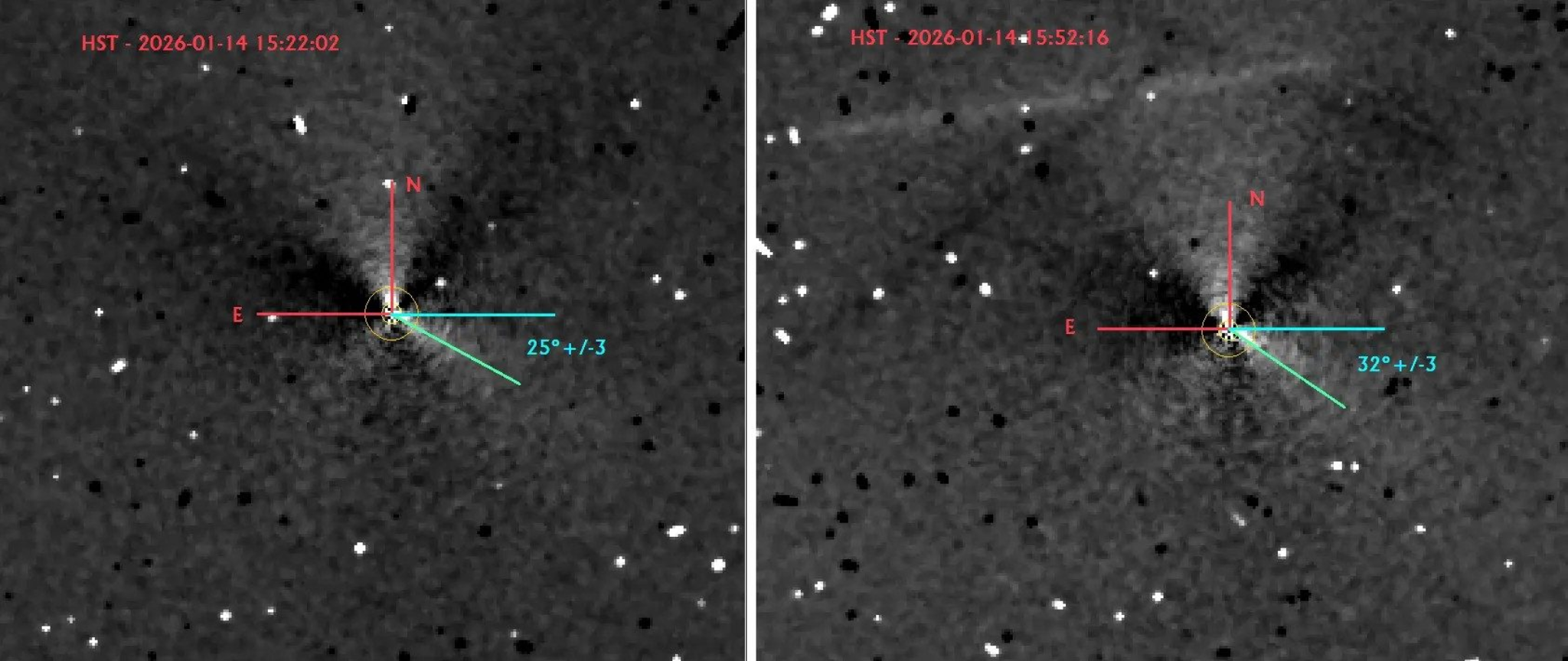}
\caption{Example of the jet position angle (PA) measurement procedure applied to HST/WFC3-UVIS Larson–Sekanina enhanced images of 3I/ATLAS obtained on 2026 January 14. The jet direction was determined by identifying the initial ridge defined by the brightest pixels emerging from the nucleus and verified by tracing the jet structure over approximately 30 pixels from the photocenter. Position angles were measured manually using an angular grid centered on the nucleus. Repeated measurements performed on different image stretches and enhancement parameters yielded consistent results within ±3°, adopted as the conservative uncertainty for all PA determinations.The January 14 image is shown only as an illustrative example of the PA measurement procedure; the PA time series used in the period analysis is reported in Table 1 and covers the HST observations obtained between 2025 November 30 and 2025 December 27.}
\label{fig:PA example measurement and error}
\end{figure*}

\begin{figure*}
\centering
\includegraphics[width=0.95\textwidth]{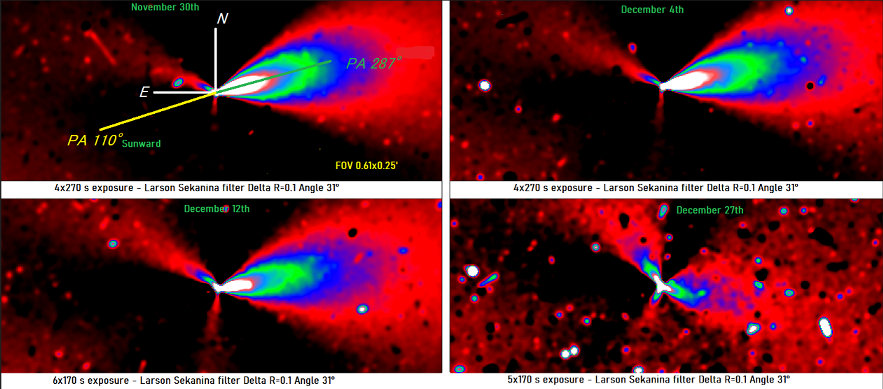}
\caption{Evolution of the post-perihelion jet morphology of 3I/ATLAS observed with HST/WFC3-UVIS on 2025 November 30, December 4, December 12, and December 27. All images were processed using the Larson–Sekanina filter (DeltaR = 0.1, rotation angle = 31°). The main jet remains visible throughout the observing campaign while exhibiting significant changes in orientation and morphology. The sunward direction is indicated in the November 30 panel. North is up and East is left.}
\label{fig:Multi epoch jets variation}
\end{figure*}

\begin{figure*}
\centering
\includegraphics[width=1.10\textwidth]{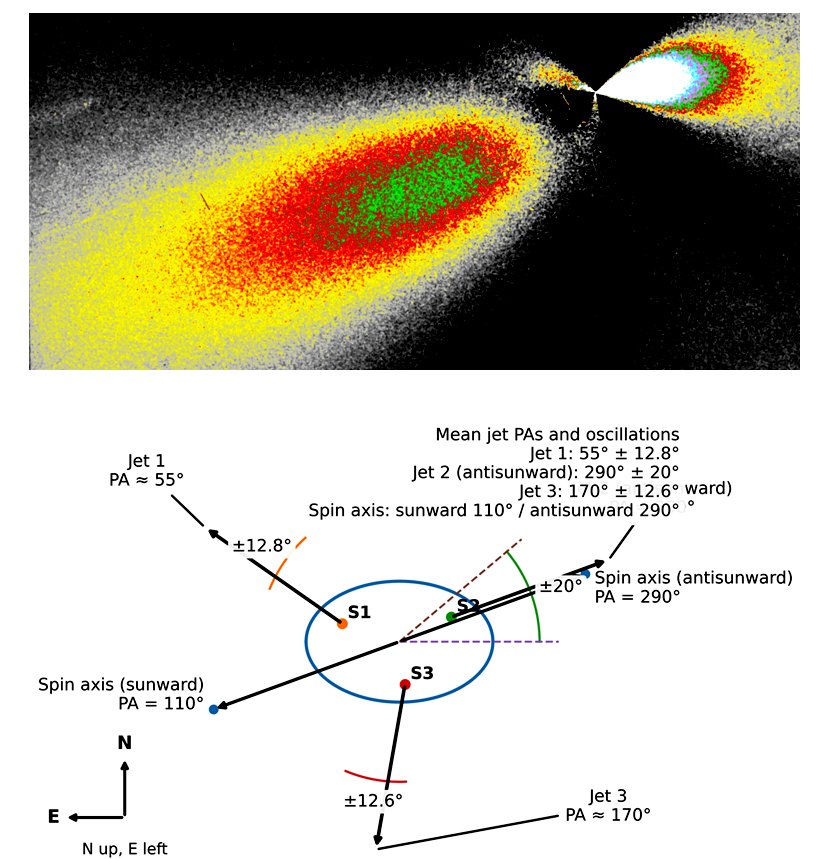}
\caption{Schematic geometry (not to scale), with North up and East to the left.
The projected spin axis is indicated in the sunward direction (PA$=110^\circ$) and anti-sunward direction (PA$=290^\circ$).
Mean jet PAs and measured PA oscillation half-amplitudes are: Jet~1, $55^\circ\pm12.8^\circ$; Jet~2 (anti-sunward), $290^\circ\pm20^\circ$; Jet~3, $170^\circ\pm12.6^\circ$.
The scheme show the anti-sunward jet in PA=290° wobbling with period of 7.2 h, that is dominant in the photometric variation also.}
\label{fig:schematic}
\end{figure*}

\subsection{\textbf{Photometry and Astrometrica apertures}}
The 3I/ATLAS photometry was performed with Astrometrica using circular apertures.
Astrometrica adopts an effective aperture radius $(N+0.5)$ pixels due to the fractional weighting of pixels at the aperture boundary radius with the sky background
estimated in the same annulus.
For a plate scale of $1.38''~{\rm px}^{-1}$, the effective radii are $r_{\rm eff}\approx6.2''$ for $N=4$ and $r_{\rm eff}\approx42.2''$ for $N=30$.
The time-series photometry used to derive the light-curve period was acquired at the Toni Scarmato’s Astronomical Observatory (observer code: MPC L92 ), using a consistent instrumental setup throughout the observing campaign (see Table 4). Observations were conducted on multiple nights under photometric or near-photometric conditions, with exposures selected to avoid saturation of field stars and to maintain adequate signal-to-noise ratio on the 3I/ATLAS while minimizing trailing.
All frames were bias/dark corrected and flat-fielded using standard CCD reduction procedures. Astrometric registration was performed to ensure consistent aperture placement across the time series. Photometric calibration used field stars from a standard catalog (Gaia DR3), and magnitudes were placed onto a common internal scale through frame-by-frame zero-point determination \textbf{(Scarmato 2025)}. The formal uncertainty on each measurement was derived from the photometric signal-to-noise ratio, and we additionally accounted for excess scatter arising from transparency
variations, background gradients from the coma, and residual systematics.
To reduce short-timescale noise and provide homogeneous sampling for the period
analysis, the time series was also analyzed in uniformly binned intervals of and 30 minutes, using inverse-variance weighted averages computed independently for each night.
These ground-based data provide an independent constraint on the rotational modulation of the inner coma brightness and enable direct comparison with the jet-derived period obtained from space-based morphology measurements. The ASI294MC Pro is a one-shot color CMOS detector, and the quoted R, G, and B measurements refer to instrumental Bayer channels rather than standard photometric passbands. In this work, the red channel is used for time-series photometry and calibrated against Gaia DR3 red-band magnitudes (denoted here as Gr in the Astrometrica output).

Photometric zero points were determined frame by frame using field stars matched to the Gaia DR3 catalog. The calibrated magnitudes are therefore placed on a Gaia-based red photometric scale, which should be regarded as an approximate transformation rather than a strict standard-system calibration.

No explicit color-term correction was applied. Since the analysis focuses on relative time-series photometry for period determination, residual passband mismatches primarily affect the absolute zero point but do not significantly influence the derived periodicity.\\
\\
\begin{deluxetable*}{lcccccc}
\tabletypesize{\footnotesize}
\tablewidth{0pt}
\tablecaption{Binned photometry of 3I/ATLAS (Dataset 2 - 30 pixels aperture); 30-point bins; $P_{\rm phot} = 7.136$ h)\label{tab:dataset2_bin30_phase7136_sem}}
\tablehead{
\colhead{UT mid-time} & \colhead{JD} & \colhead{Phase at $P_{\rm phot}$} & \colhead{$m^\prime$ (mag)} & \colhead{$\sigma_{m^\prime}$ (mag)} & \colhead{$N$}
}
\startdata
2025-12-09 01:25:00 & 2461018.559025 & 0.000 & 12.137 & 0.024 & 30 \\
2025-12-09 02:04:11 & 2461018.586243 & 0.092 & 12.043 & 0.025 & 30 \\
2025-12-09 02:46:01 & 2461018.615292 & 0.189 & 12.185 & 0.023 & 30 \\
2025-12-09 03:06:45 & 2461018.629684 & 0.238 & 12.212 & \nodata & 1 \\
2025-12-15 02:13:53 & 2461024.592972 & 0.294 & 11.157 & 0.018 & 30 \\
2025-12-15 02:44:03 & 2461024.613921 & 0.364 & 11.270 & 0.019 & 28 \\
2025-12-18 02:54:57 & 2461027.621498 & 0.479 & 11.504 & 0.028 & 30 \\
2025-12-18 03:25:34 & 2461027.642760 & 0.551 & 11.618 & 0.020 & 30 \\
2025-12-18 03:49:33 & 2461027.659410 & 0.607 & 11.577 & 0.032 & 17 \\
2025-12-22 22:21:49 & 2461032.431815 & 0.657 & 11.538 & 0.029 & 30 \\
2025-12-22 22:52:34 & 2461032.453175 & 0.729 & 11.506 & 0.045 & 29 \\
2025-12-27 22:24:23 & 2461037.433599 & 0.480 & 11.519 & 0.033 & 30 \\
2025-12-27 23:15:56 & 2461037.469394 & 0.600 & 11.675 & 0.025 & 30 \\
2025-12-27 23:53:08 & 2461037.495228 & 0.687 & 11.894 & 0.033 & 30 \\
2025-12-28 00:20:53 & 2461037.514498 & 0.752 & 11.856 & 0.032 & 23 \\
\enddata
\footnote{}{Magnitudes $m^\prime$ are corrected to a common photometric scale using the median zero-point of 27.397 mag for Dataset 2. Phases are computed for $P_{\rm phot} = 7.142$ h, with the earliest binned point in this dataset taken as phase zero. Each row corresponds to a bin of 30 consecutive measurements within a single observing night; the final bin of a night may contain fewer than 30 points. The binned magnitude is computed from the arithmetic mean flux in each bin. The quoted uncertainty $\sigma_{m^\prime}$ is the error on the binned point, computed as the standard error of the mean of the corrected magnitudes, $\sigma_{m^\prime} = s/\sqrt{N}$. All measurements use filter Gr R.}
\end{deluxetable*}

\begin{deluxetable*}{lcccccc}
\tabletypesize{\footnotesize}
\tablewidth{0pt}
\tablecaption{Binned photometry of 3I/ATLAS (Dataset 1 - 4 pixels aperture); 30-point bins; $P_{\rm phot} = 7.136$ h)\label{tab:dataset1_bin30_phase7136_sem}}
\tablehead{
\colhead{UT mid-time} & \colhead{JD} & \colhead{Phase at $P_{\rm phot}$} & \colhead{$m^\prime$ (mag)} & \colhead{$\sigma_{m^\prime}$ (mag)} & \colhead{$N$}
}
\startdata
2025-12-09 01:23:48 & 2461018.558197 & 0.000 & 14.577 & 0.029 & 30 \\
2025-12-09 02:02:45 & 2461018.585248 & 0.091 & 14.339 & 0.021 & 30 \\
2025-12-09 02:44:17 & 2461018.614086 & 0.188 & 14.577 & 0.027 & 30 \\
2025-12-09 03:05:23 & 2461018.628738 & 0.237 & 14.601 & 0.083 & 3 \\
2025-12-15 02:14:42 & 2461024.593538 & 0.298 & 13.657 & 0.022 & 30 \\
2025-12-15 02:44:35 & 2461024.614291 & 0.368 & 13.649 & 0.051 & 27 \\
2025-12-18 02:54:57 & 2461027.621498 & 0.482 & 13.957 & 0.030 & 30 \\
2025-12-18 03:25:34 & 2461027.642760 & 0.553 & 14.110 & 0.023 & 30 \\
2025-12-18 03:49:33 & 2461027.659410 & 0.609 & 14.027 & 0.026 & 17 \\
2025-12-22 22:21:49 & 2461032.431815 & 0.660 & 13.881 & 0.083 & 30 \\
2025-12-22 22:52:34 & 2461032.453175 & 0.732 & 14.031 & 0.038 & 29 \\
2025-12-27 22:24:23 & 2461037.433599 & 0.482 & 14.097 & 0.029 & 30 \\
2025-12-27 23:19:03 & 2461037.471558 & 0.610 & 14.121 & 0.057 & 30 \\
2025-12-28 01:09:06 & 2461037.547991 & 0.867 & 13.715 & 0.051 & 24 \\
\enddata
\footnote{}{Magnitudes $m^\prime$ are corrected to a common photometric scale using the median zero-point of 27.397 mag for Dataset 1. Phases are computed for $P_{\rm phot} = 7.136$ h, with the earliest binned point in this dataset taken as phase zero. Each row corresponds to a bin of 30 consecutive measurements within a single observing night; the final bin of a night may contain fewer than 30 points. The binned magnitude is computed from the arithmetic mean flux in each bin. The quoted uncertainty $\sigma_{m^\prime}$ is the error on the binned point, computed as the standard error of the mean of the corrected magnitudes, $\sigma_{m^\prime} = s/\sqrt{N}$. All measurements use filter Gr R.}
\end{deluxetable*}

\begin{deluxetable}{ll}
\tablecaption{Core technical specifications of the cooled one-shot color ZWO ASI294MC Pro CMOS camera\label{tab:asi294_specs} and TELESCOPE SETUP}
\tablehead{\colhead{Parameter} & \colhead{Value}}
\tabletypesize{\small}
\startdata
Sensor & Sony IMX294 CMOS, OSC, back-illuminated \\
Bands & R (580 nm peak), G (530 nm peak), B (470 nm peak) \\
Sensor format & 4/3 in; 23.2 mm diagonal \\
Resolution & 11.7 MP; 4144 $\times$ 2822 pixels \\
Pixel size & 4.63 $\mu$m (9.26 $\mu$m BIN 2x2)\\
Active area & 19.2 $\times$ 13.0 mm \\
ADC resolution & 14 bit \\
Dynamic range & 13 stops \\
Full well capacity & 63.7 ke$^{-}$ \\
Read noise & 1.2-7.3 e$^{-}$ \\
Peak quantum efficiency & about 75\% \\
Shutter type & Rolling shutter \\
Exposure range & 32 $\mu$s to 2000 s \\
Frame rate & 19 fps (12-bit high speed); 16 fps (14-bit) \\
Computer interface & USB 3.0 with integrated USB 2.0 hub \\
Internal buffer & 256 MB DDR3 \\
Mechanical thread & M42 $\times$ 0.75 \\
Back focus & 6.5 mm \\
Protective window & Built-in IR-cut window \\
Power requirement & 12 V DC; up to 5 A max (stable) \\
Cooling system & Regulated two-stage Peltier TEC cooling \\
Cooling performance & About 35-40 $^{\circ}$C below ambient \\
TELESCOPE & NEWTON reflector \\
Aperture & 0.25 m \\
Focal length & 1200 cm (1380 cm with coma corrector) \\
scale & 0.74"/pixel (1.38"/pixel bin 2x2 with coma corrector) \\
FOV & 41'x37' \\
\enddata
\end{deluxetable}

\begin{deluxetable}{ll}
\tablecaption{Main characteristics of the HST WFC3/UVIS channel}
\tablehead{
\colhead{Parameter} & \colhead{Value}
}
\startdata
Instrument & Wide Field Camera 3 (WFC3) \\
Channel & UVIS (Ultraviolet-Visible) \\
Detector type & CCD (2 $\times$ e2v CCDs) \\
Array size & $2 \times (2051 \times 4096)$ pixels \\
Field of view & $\sim 162'' \times 162''$ \\
Pixel scale & $0.0395''$/pixel \\
Pixel size & 15 $\mu$m \\
Wavelength range & $\sim 200$--1000 nm \\
Quantum efficiency (peak) & $\sim 40$--50\% (visible) \\
Read noise & $\sim 3$ e$^-$ \\
Gain & 1.5--2.5 e$^-$/DN \\
Full well capacity & $\sim 70{,}000$ e$^-$ \\
Dark current & $\sim 5 \times 10^{-4}$ e$^-$/s/pixel \\
ADC resolution & 16 bit \\
Shutter & Mechanical \\
Typical PSF (FWHM) & $\sim 0.07''$ ($\sim$2 pixels) \\
Filters & Broad, medium, narrow band (UV--optical) \\
Data products & FLT, FLC, DRZ, DRC \\
\enddata

\end{deluxetable}
\begin{deluxetable}{lc}
\tablecaption{Summary of the independent photometric period analysis (30-minute bins for 4 pixels aperture).\label{tab:phot}}
\tablehead{\colhead{Quantity} & \colhead{Value}}
\startdata
Band / binning & Gr ($R$ band) / 30 min \\
N nights / N bins & 4 / 14 \\
Best period $P_{\rm phot}$ & 7.136 h \\
$1\sigma$ on $P_{\rm phot}$ & 0.001 h \\
Semi-amplitude $A$ & 0.311 mag \\
Peak-to-peak $2A$ & 0.622 mag \\
Jitter $\sigma_{\rm jit}$ & 0.089 mag \\
\enddata
\end{deluxetable}
\
\section{Methods}
\subsection{Harmonic representation of PA oscillations}
For each trial period P, we computed the rotational phase of each jet-PA measurement as
$\phi_i(P) = frac[(t_i - t_0)/P]$,
where $t_i$ is the observation time and $t_0$ is an arbitrary reference epoch. The phase dependence of the Jets PA was modeled with a truncated Fourier series, with $K = 1$ and $K = 2$. At fixed P, the coefficients were obtained by weighted least squares, adopting a conservative uncertainty $\sigma(PA)$ = 3 degs for each measurement and weights $w_i = 1/\sigma^2(PA)$. We scan P on a uniform grid in the physically relevant window and record $ \chi ^ 2 (P)$ for each model order. The preferred jet period was taken as the minimum of $\chi ^ 2 (P)$, with K = 2 adopted because it better reproduces the non-sinusoidal waveform of the phase-folded PA curve.
Because the PA measurements are sparse in time and clustered within a limited number of epochs, and because the measured PA depends to some extent on morphology-dependent centroiding of the brightest jet feature, the uncertainty on $P_jet$ is not interpreted as a purely formal statistical error. Instead, the quoted uncertainty reflects a conservative estimate of the range of periods that provide comparably acceptable fits once sparse sampling and morphology-dependent systematics are taken into account. We therefore quote $P_jet$ = 7.20 ± 0.05 h as a conservative interval rather than as a formal $\Delta \chi^2 = 1$ confidence limit.
We describe the PA time series with a k-harmonic representation,
\begin{equation}
{\rm PA}(t) = {\rm PA}_0 + \sum_{k=1}^{k} A_k \sin\!\left(\frac{2\pi k}{P_{\rm prec}}t+\phi_k\right),
\end{equation}
where $P_{\rm prec}$ is interpreted as an attitude precession period.
We use $k=1$ and $k=2$.

\subsection{\textbf{Photometric folding and stationary tests}}
Coma photometry is folded at trial periods near $P_{\rm prec} \sim 7.1$ h. We assess variability by comparing four-night and five-night folds, allowing per-night offsets. For the large-aperture data, where morphology-dependent scatter dominates over the photometric uncertainties, we use an unweighted least-squares / residual-variance period scan rather than a formally weighted $\chi^2$ statistic.\\
\
\subsubsection{\textbf{Zero-point normalization}}
When individual frame zero-points $ZP_i$ were available, we homogenized the
photometry to a common reference zero-point $ZP_{\rm ref}$ (taken as the median
over the dataset) as
\begin{equation}
m_{{\rm corr},i} = m_i + \left(ZP_{\rm ref} - ZP_i\right).
\label{eq:zp_correction}
\end{equation}
\textbf{To assess the stability of the photometric calibration, we analyzed the frame-by-frame zero-point variations for each dataset. The resulting distributions are shown in Figure 4, where no significant systematic trends are observed within individual nights. This supports the use of a common reference zero point for the normalization procedure.}

\subsection{\textbf{Zero-point stability}}
To verify the stability of the photometric calibration, we examined the frame-by-frame zero points reported in Tables 9 and 10 for the 4-pixel and 30-pixel apertures, respectively. Figure 4 shows the residual zero-point variations relative to the median value of each dataset.
For the 4-pixel aperture, the median zero point is 27.693 mag with a standard deviation of 0.226 mag. For the 30-pixel aperture, the median zero point is 27.421 mag with a standard deviation of 0.324 mag. The larger scatter observed for the 30-pixel aperture is expected because the larger photometric aperture includes a greater fraction of the diffuse coma and sky background.
No monotonic temporal drift is observed in either dataset. The measured frame-by-frame zero points were therefore used directly in the photometric normalization procedure described in Eq. (3), ensuring that variations in atmospheric transparency and calibration do not introduce spurious periodic signals into the time-series analysis. 

\begin{figure*}
\centering
\includegraphics[width=0.95\textwidth]{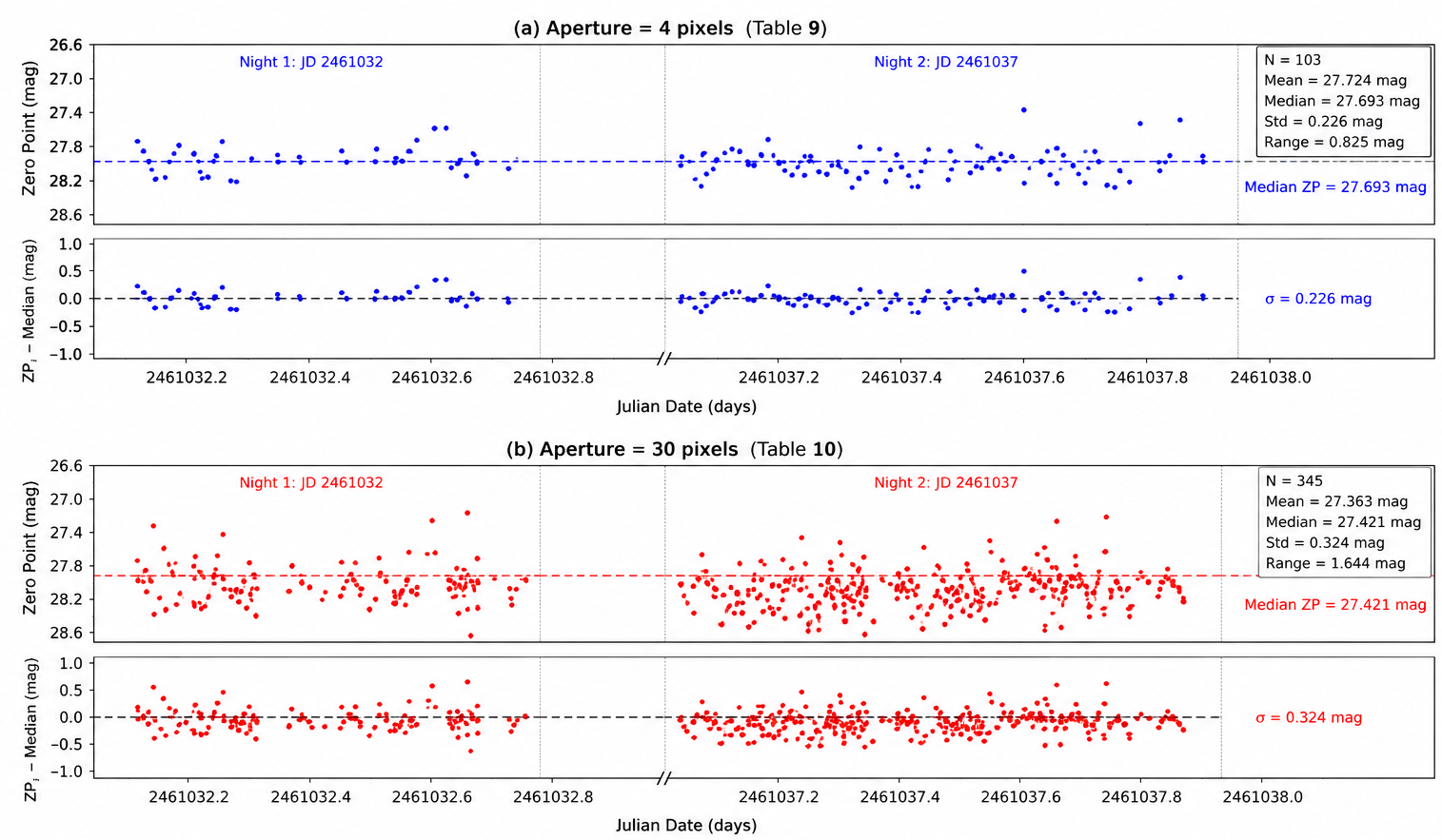}
\caption{Frame-by-frame photometric zero-point stability for the two Astrometrica apertures used in the time-series analysis. Panel (a) shows the results for the 4-pixel aperture (Table 9), while panel (b) shows the results for the 30-pixel aperture (Table 10). In each panel, the upper subpanel displays the measured zero point as a function of Julian Date, whereas the lower subpanel shows the residuals relative to the median value of the corresponding dataset, ($ZP_i-\widetilde{ZP}$). The dashed horizontal lines indicate the median zero points (($\widetilde{ZP}=27.693$) mag for the 4-pixel aperture and ($\widetilde{ZP}=27.421$) mag for the 30-pixel aperture). No significant monotonic drift is observed in either dataset. The larger scatter observed for the 30-pixel aperture is expected because the larger aperture includes a greater contribution from diffuse coma emission and sky background. These results demonstrate that the photometric normalization based on frame-by-frame zero-point corrections effectively removes calibration variations and cannot account for the periodic signals detected in the photometric time series.}
\label{fig:Zero-Point analisys}
\end{figure*}

\subsection{\textbf{Seeing Dependence of the Photometric Residuals}}
{To evaluate whether atmospheric image quality could influence the derived periodic signal, we examined the dependence of the photometric residuals on the measured seeing values for both photometric apertures. The seeing ranged between approximately 5.0–6.0 arcsec for the 4-pixel aperture dataset and between 3.3–4.1 arcsec for the 30-pixel aperture dataset (Table 11 and Table 12). No systematic trend between residual magnitude and seeing was detected. Frames obtained under poorer seeing conditions do not show larger residuals than frames obtained under better seeing conditions. This indicates that the observed photometric modulation is not driven by seeing fluctuations but reflects intrinsic variability of the coma and jet activity. The absence of a significant correlation supports the robustness of the derived periodic signal.(see Figure 5)}

\begin{figure*}
\centering
\includegraphics[width=0.95\textwidth]{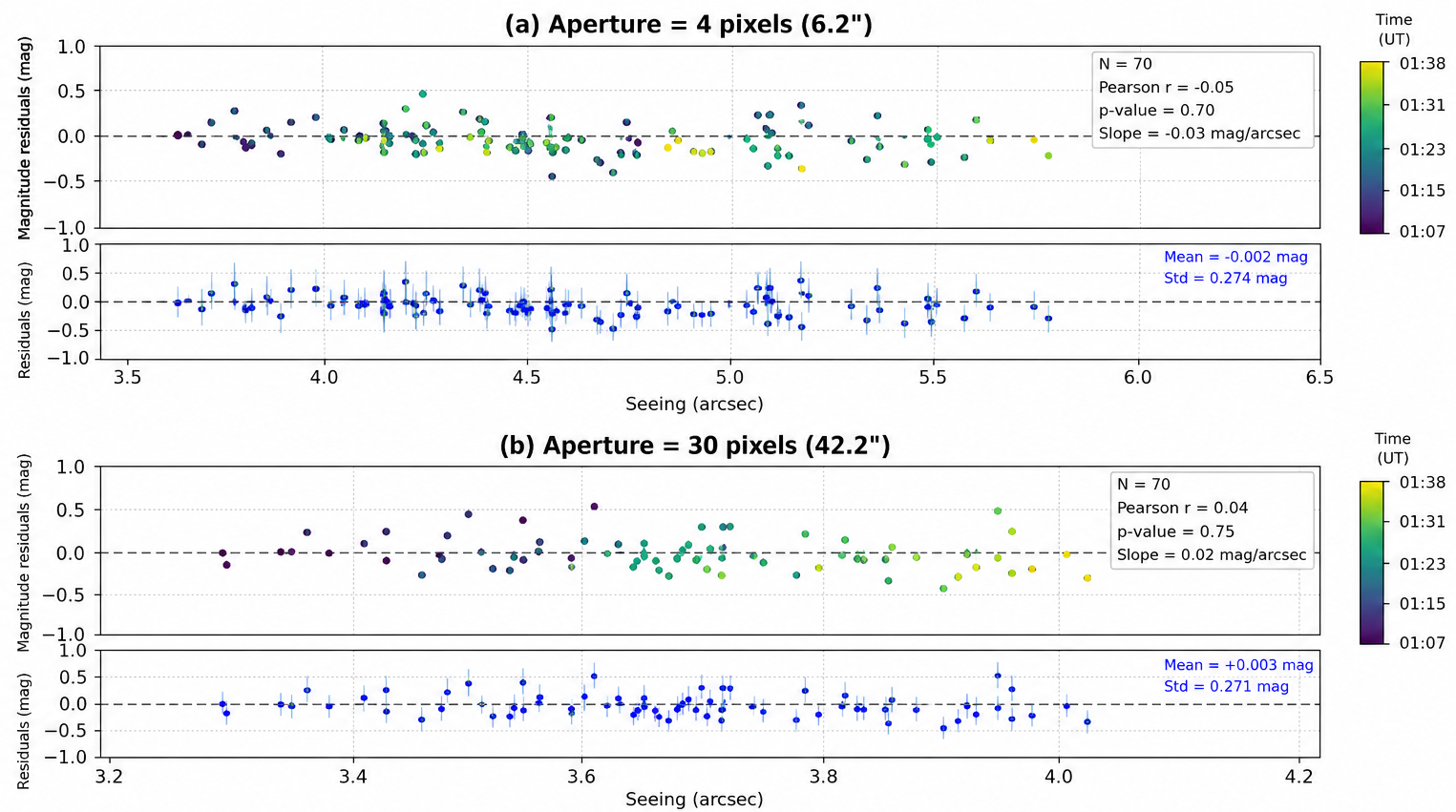}
\caption{Dependence of the photometric residuals on atmospheric seeing for the two photometric apertures used in the analysis. Panel (a) shows the 4-pixel aperture (6.2''), while panel (b) shows the 30-pixel aperture (42.2''). In each panel, the residual magnitudes are plotted as a function of the measured seeing. The dashed horizontal line marks zero residual. No significant correlation is observed between residual magnitude and seeing for either aperture, indicating that variations in atmospheric image quality do not account for the detected photometric modulation. The absence of a systematic trend supports the interpretation that the observed periodic signal is intrinsic to the cometary coma and jet activity rather than an artifact of seeing fluctuations..}
\label{fig:Seeing vs magnitudes}
\end{figure*}

\subsubsection{\textbf{Photometric uncertainties and weights}}
Formal magnitude uncertainties were estimated from the reported signal-to-noise ratio ${\rm SNR}_i$ as
\begin{equation}
\sigma_i = \frac{1.0857}{{\rm SNR}_i},
\label{eq:sigma_snr}
\end{equation}
and used as inverse-variance weights $w_i = \sigma_i^{-2}$ in a weighted least
squares (WLS) fit.\\

\subsubsection{\textbf{Within-night binning}}
To reduce short-timescale scatter, data were optionally binned within each
night using a time bin $\Delta t$ (e.g., 5, 10, or 30 minutes). For a bin
containing measurements $j$, the weighted binned time and magnitude were
computed as

\begin{equation}
t_{\rm bin} = \frac{\sum_j w_j t_j}{\sum_j w_j}\\
\end{equation}
\begin{equation}m_{\rm bin} = \frac{\sum_j w_j m_{{\rm corr},j}}{\sum_j w_j}\\
\end{equation}
\begin{equation}
\sigma_{\rm bin} = \left(\sum_j w_j\right)^{-1/2}
\end{equation}
\
\subsubsection{\textbf{Single-harmonic model with nightly offsets}}
At a trial period $P$ (frequency $f=1/P$; angular frequency
$\omega = 2\pi/P$), we fitted the binned (or unbinned) magnitudes with
\begin{equation}
m_i = C + O_{k(i)} + a \cos(\omega t_i) + b \sin(\omega t_i),
\label{eq:model_1harm}
\end{equation}
where $C$ is a global constant, $O_{k(i)}$ is the additive offset for the night
$k$ to which point $i$ belongs, and $a,b$ are the harmonic coefficients. To fix
the reference level we set
\begin{equation}
O_{1} = 0,
\label{eq:offset_constraint}
\end{equation}
so that all other offsets are measured relative to the first night. The
semi-amplitude of the modulation is
\begin{equation}
A = \sqrt{a^2 + b^2},
\qquad
\Delta m_{\rm p2p} = 2A.
\label{eq:amplitude}
\end{equation}

\subsubsection{\textbf{Objective function and model selection}}
For each trial period $P$, the best-fitting parameters were obtained by WLS
minimization of
\begin{equation}
\chi^2(P) = \sum_i \left(\frac{m_i - m_{\rm model}(t_i;P)}{\sigma_i}\right)^2,
\label{eq:chi2}
\end{equation}
with ${\rm dof} = N - k$ degrees of freedom, where $N$ is the number of points
and $k$ the number of fitted parameters. We also report the Bayesian
Information Criterion
\begin{equation}
{\rm BIC}(P) = \chi^2(P) + k \ln N.
\label{eq:bic}
\end{equation}

\subsubsection{\textbf{Period search and oversampling}}
We searched periods in the range $P_{\min} \le P \le P_{\max}$ by scanning a
uniform frequency grid in cycles per day. For a total time baseline
$T_{\rm span} = t_{\max} - t_{\min}$, the frequency step was set to
\begin{equation}
\Delta f = \frac{1}{{\rm OS}\,T_{\rm span}},
\label{eq:dfreq}
\end{equation}
where ${\rm OS}$ is an oversampling factor. The best period $P_{\rm best}$ was
taken as the minimum of $\chi^2(P)$; we refined the solution with a denser
local grid around the global minimum. A formal $1\sigma$ uncertainty on $P$ was
estimated from the $\Delta\chi^2=1$ criterion:
\begin{equation}
\chi^2(P) = \chi^2_{\min} + 1.
\label{eq:dchi2}
\end{equation}

\subsubsection{\textbf{Additional scatter (jitter)}}
Because SNR uncertainties account primarily for statistical noise, we
estimated an additional ``jitter'' term $\sigma_{\rm jit}$ such that
\begin{equation}
\sigma_{{\rm eff},i}^2 = \sigma_i^2 + \sigma_{\rm jit}^2,
\label{eq:sigma_eff}
\end{equation}
and $\sigma_{\rm jit}$ was chosen to agree approximately with the number of degrees of freedom (dof),
\begin{equation}
\sum_i \left(\frac{r_i}{\sigma_{{\rm eff},i}}\right)^2 \simeq {\rm dof},
\label{eq:jitter_condition}
\end{equation}
where $r_i$ are the residuals of the best-fitting model.

\subsubsection{\textbf{Phase folding (0--1 cycles) and sinusoidal representation}}
For visualization we removed the fitted nightly offsets and folded the data on
$P_{\rm best}$. The offset-corrected magnitudes are
\begin{equation}
\Delta m_i = m_i - \left(C + O_{k(i)}\right),
\label{eq:delta_mag}
\end{equation}
and the rotational phase in the interval $[0,1)$ is defined as
\begin{equation}
\phi_i = {\rm frac}\!\left(\frac{t_i - t_0}{P_{\rm best}}\right),
\qquad 0 \le \phi_i < 1,
\label{eq:phase_def}
\end{equation}
where $t_0$ is an arbitrary reference epoch (e.g., the first observation) and
${\rm frac}(x)=x-\lfloor x \rfloor$ denotes the fractional part. The
corresponding best-fitting sinusoid (in $\Delta m$) is
\begin{equation}
\Delta m_{\rm model}(\phi) = a \cos(2\pi \phi) + b \sin(2\pi \phi).
\label{eq:folded_model}
\end{equation}

\subsubsection{\textbf{Approach for small- and large-aperture photometry}} For the small-aperture photometry (4-pixel radius), where statistical uncertainties dominate and the light-curve morphology remains stable, we adopt a weighted least-squares approach based on formal measurement uncertainties.

However, for the large-aperture photometry (30-pixel radius), the scatter is dominated by morphology-dependent systematic effects, including evolving coma structures and feature-dependent brightness variations. In this regime, weighting by formal photometric uncertainties would artificially overemphasize small statistical errors and underestimate the true variability.

Therefore, for the large-aperture dataset, we adopt an unweighted least-squares approach and interpret the periodograms in terms of residual variance rather than formal $\chi^2$ statistics. This approach follows standard practice when systematic variability dominates over formal measurement uncertainties.
This distinction ensures statistical consistency across datasets with different systematic properties.
\\
\section{Results}
\subsection{\textbf{PA wobble period from HST morphology and photometric variations}}
Using HST images processed through a Larson--Sekanina Rotational-Gradient Filter (with typical settings of $\delta r=0.1$ and $\Delta\theta=31^\circ$), we measured the PAs of the dominant jet-like structure across epochs spanning 2025 November 30th--December 27th.
A weighted period scan with a truncated Fourier representation (up to $k=2$ harmonics) yields a best-fitting wobble period
\begin{equation}
P_{\rm jet}=7.20\pm0.05~{\rm h},
\end{equation}
where the uncertainty is dominated by sparse sampling and morphology-dependent systematics (we adopt $\sigma_{\rm PA}=3^\circ$ per measurement).
The associated PA excursion is of order $\sim20^\circ$ and the inferred projected axis is aligned with the sunward direction to within $\sim20^\circ$.
Ground-based time-series photometry (MPC station code L92; $G_r$ band $\approx R$), was extracted using Astrometrica with an aperture of $N=4$ pixels (effective radius $r_{\rm eff}\approx (N+0.5)\times 1.38''\approx6.2''$) shows a coherent modulation over a four-night subset after allowing nightly offsets and 30-minute binning.
A Fourier-series period scan returns
\begin{equation}
P_{\rm phot}=7.136\pm0.001~{\rm h}
\end{equation}
with a formal $1\sigma$ uncertainty, semi-amplitude $A\simeq0.311$~mag, and additional scatter $\sigma_{\rm jit}\simeq0.089$~mag.
Because the measured flux is coma-dominated, we treat the formal uncertainty as a lower bound: systematic effects (seeing, aperture losses, morphology changes) can dominate over photon noise.
We repeated the photometric analysis with a larger Astrometrica aperture of $N = 30$ pixels (effective radius $r_{\rm eff} \approx 42.2''$), designed to capture a broader coma region that is sensitive to fan-like jet opening.

For this large-aperture data set, the formal point-by-point photometric uncertainties are much smaller than the actual scatter around the Fourier models. We therefore do not interpret the period scan as a formally weighted $\chi^2$ statistic. Instead, after allowing per-night offsets, we use an unweighted least-squares scan and evaluate the trial periods through the residual variance (equivalently, the sum of squared residuals). In this sense, the periodograms shown in Figures 2 and 3 should be interpreted as residual-variance scans rather than formal $\Delta\chi^2$ periodograms.

In the physically motivated window 6.8--7.4 h, the preferred periods are:
\begin{itemize}
\item \textit{four-night subset} ($N = 285$ points): $P_{\rm best}(k=1)=7.142$ h and $P_{\rm local}(k=2)=7.179$ h, with RMS $\approx 0.147$ mag (Figure 2);
\item \textit{five-night full set} ($N = 398$ points): $P_{\rm best}(k=1)=7.158$ h and $P_{\rm local}(k=2)=7.175$ h, with RMS decreasing from $\approx 0.161$ mag ($k=1$) to $\approx 0.156$ mag ($k=2$) (Figure 3).
\end{itemize}

Including December 27, 2025 degrades phase stationarity relative to the four-night subset, consistent with the appearance of fan-shaped jets and a changing jet-to-photometry transfer function (Figure 6). Thus, the large-aperture photometry still supports a characteristic period near $\sim 7.1$ h, but this evidence should be interpreted in terms of residual-variance minima under substantial morphology-dependent scatter, rather than as a formal weighted-$\chi^2$ detection.
\textbf{The photometric measurements are significantly smaller than the scatter induced coma morphology and variable activity. We therefore adopt an unweighted least-squares approach, appropriate when systematic scatter dominates over formal measurement uncertainties . In this regime, weighting by formal measurement errors would artificially overemphasize small statistical uncertainties and underestimate systematic variations. Therefore, all period scans and Fourier fits presented in this work are computed using unweighted data, consistent with the recommendations of Scargle (1982) and subsequent astrostatistical studies. The period search was performed through a dense grid scan over the trial period, with linear least squares optimization of the harmonic coefficients at each step. For each period of trial P, the data were phased and fitted using a Fourier series of order k. (See Section 3.2). We adopted $k<=2$ to allow for non-sinusoidal light-curve shapes while avoiding overfitting. For each period of trial, the residual variance was calculated and the best period was identified as the global minimum of the residual variance. The uncertainty on the period was estimated from the width of the minimum in the periodogram, combined with sampling limitations and morphology-dependent systematics. o verify that the observed photometric modulation is not driven by image quality variations, we compared both the corrected magnitudes and the residuals of the best-fitting model with the measured seeing (FWHM). No statistically significant correlation was found (Figure 5), supporting the interpretation that the variability is intrinsic to the cometary activity rather than seeing-induced.}

\begin{figure*}
\centering
\includegraphics[width=0.95\textwidth]{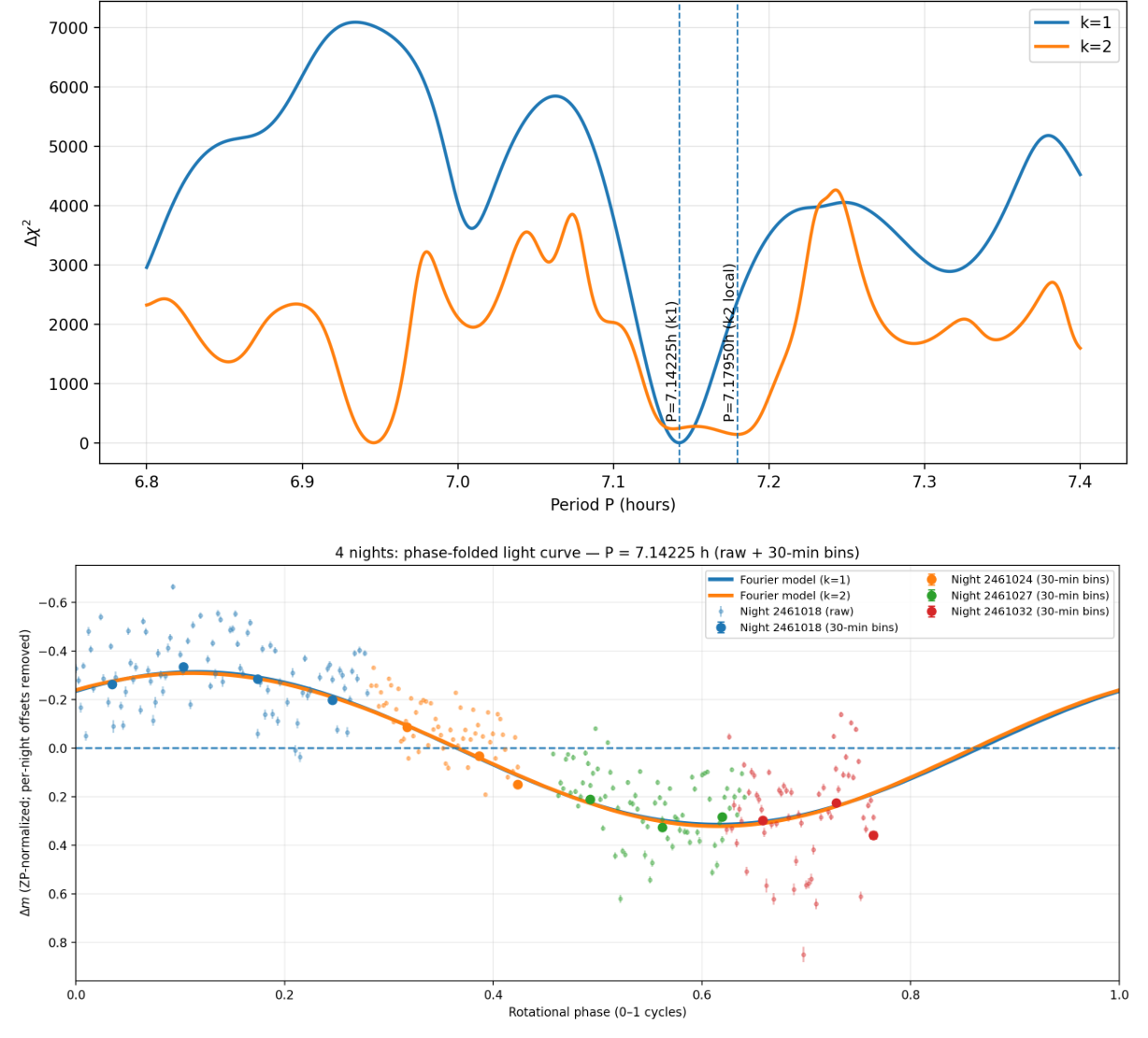}
\caption{Large-aperture ($N = 30$, $r_{\rm eff} \approx 42.2''$) photometry: four-night subset. Top: unweighted residual-variance period scan over 6.8--7.4 h for $k = 1$ and $k = 2$ Fourier models (per-night offsets allowed). Because the formal photometric measurement errors are much smaller than the total scatter, the scan is presented as a variance-reduction / residual-variance diagnostic rather than a formal $\Delta\chi^2$ statistic. Bottom: phase-folded light curve at $P = 7.142$ h after removing per-night offsets; the $k = 2$ model refines the waveform shape while preserving inter-night phase coherence. The plotted error bars represent only formal measurement uncertainties and should be regarded as lower bounds to the true scatter.}
\label{fig:phot4n}
\end{figure*}

\begin{figure*}
\centering
\includegraphics[width=0.87\textwidth]{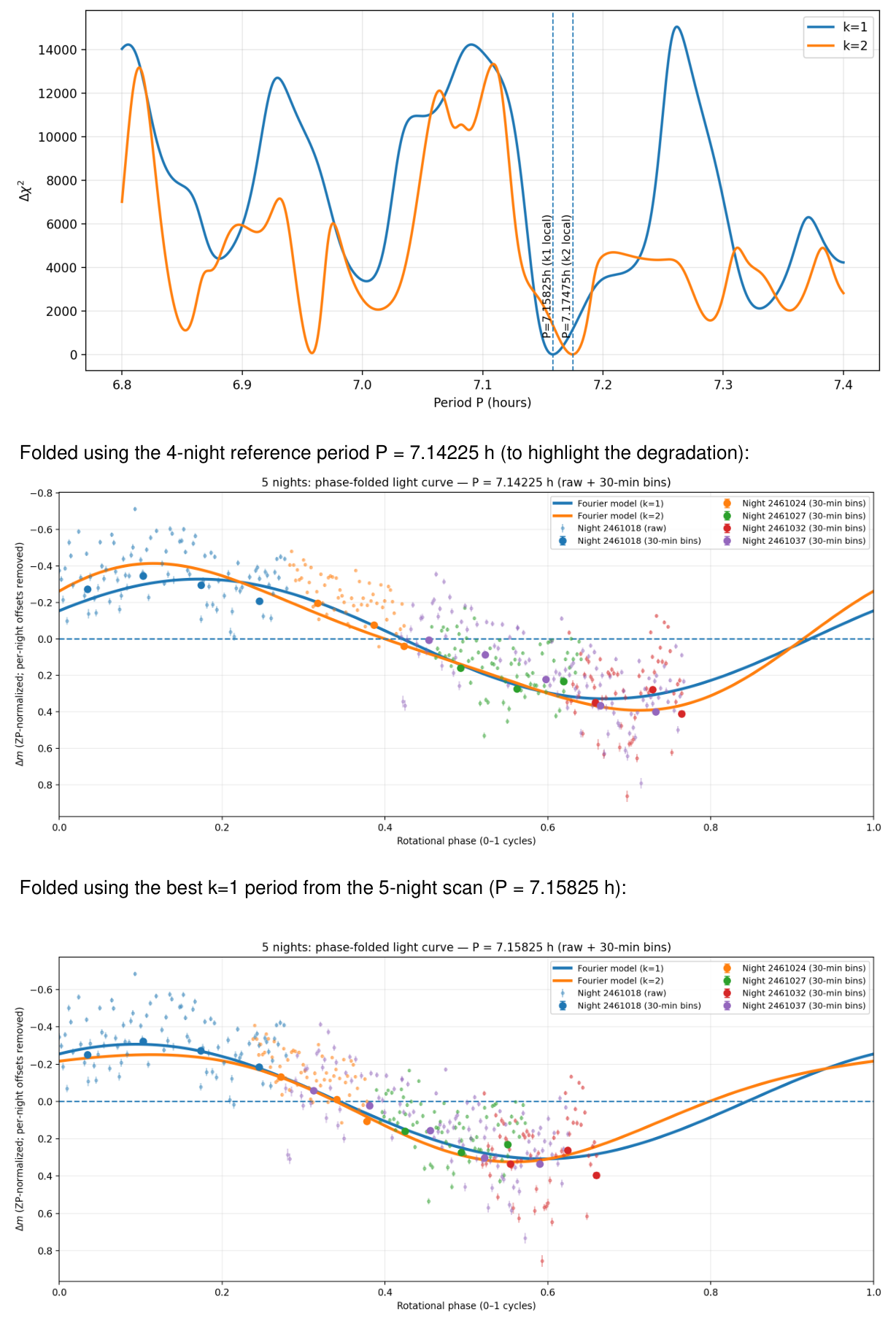}
\caption{Same as Figure 2, but for the full five-night data set including December 27, 2025. Top: unweighted residual-variance period scan within 6.8--7.4 h. Middle: fold at the four-night reference period $P = 7.142$ h, highlighting degraded stationarity introduced by the fifth night. Bottom: fold at the best $k = 1$ period from the five-night scan, $P = 7.158$ h.}
\label{fig:phot5n}
\end{figure*}

\begin{figure*}
\centering
\includegraphics[width=0.95\textwidth]{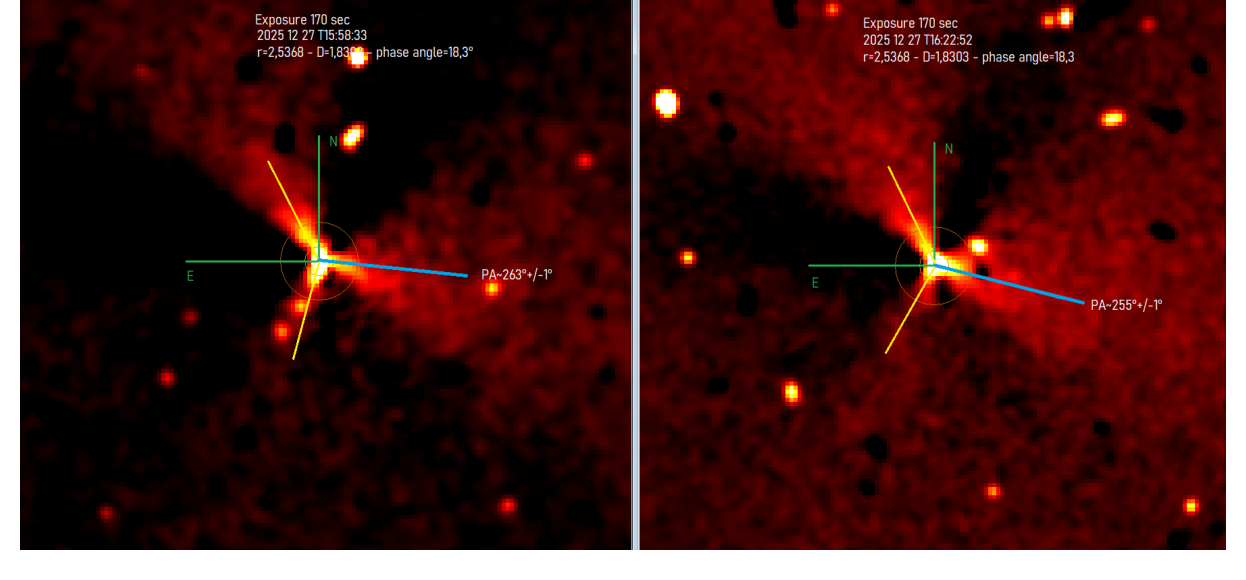}
\caption{Inner-coma jets of 3I/ATLAS on 2025-12-27 UT. Two principal jet structures remain separated by $\sim120^\circ$,
but the emission broadens into a fan. Such morphological changes modulate the coma brightness within a fixed aperture and can alter the amplitude/shape (and sometimes the apparent phase) of the photometric signal.}
\label{fig:jets20251227}
\end{figure*}

\subsection{\textbf{Minor jets: additional PA periodicity from Jet 1 and Jet 3}}
Independent PA time series for the secondary and tertiary jet structure yield shorter best-fit periods from weighted sinusoid$+$trend fits of the form
${\rm PA}(t)=c_0+c_1 t + a\sin(2\pi t/P)+b\cos(2\pi t/P)$, where the linear term captures slow baseline drifts across nights.

\begin{figure*}
\centering
\includegraphics[width=0.80\textwidth]{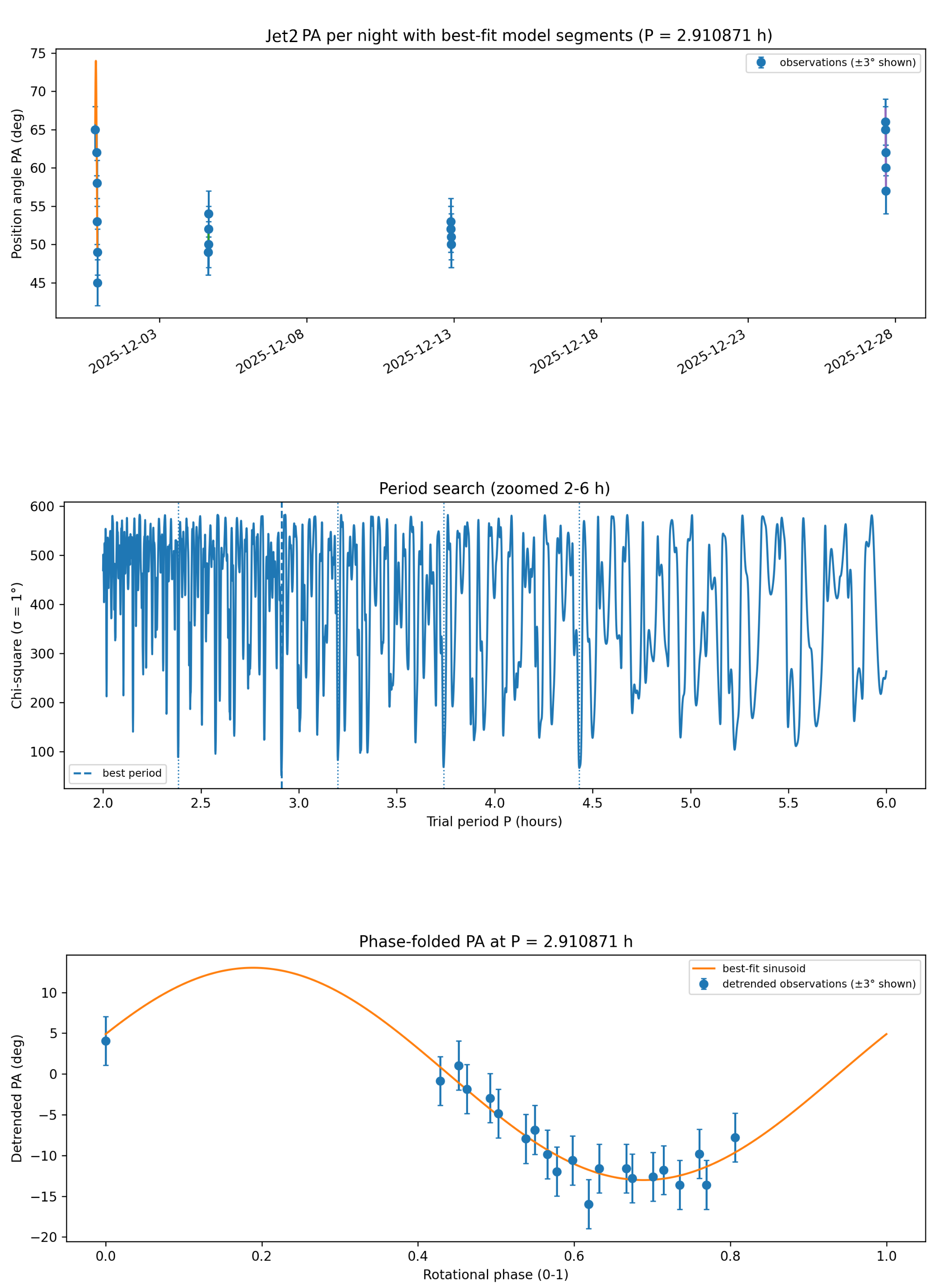}
\caption{Jet~1 diagnostic plots from the PA time series analysis: (top) PA versus time by observing segment with the best-fit model drawn within each night; (middle) period scan (2--6~h zoom); (bottom) detrended phase fold at $P_2=2.9$~h. Error bars correspond to the reported $\pm3^\circ$ values.}
\label{fig:jet2diag}
\end{figure*}

\paragraph{Jet 1.}
\textbf{For Jet~1, the best period is $P_2 = 2.911$~h with amplitude $A_2=12.8^\circ$ and a slow drift $c_1=0.009~{\rm deg~h^{-1}}$ ($\approx0.22~{\rm deg~day^{-1}}$).
Assuming the reported $\pm3^\circ$ values represent $3\sigma$ (so $\sigma=1^\circ$), the reduced $\chi^2$ at the best period is $\chi^2_\nu\simeq3.0$, indicating additional systematics.
Strong alias minima occur at $P\approx 4.432$~h, 3.739~h, 3.199~h, and 2.384~h (see Figure~\ref{fig:jet2diag}).}

\begin{figure*}
\centering
\includegraphics[width=0.80\textwidth]{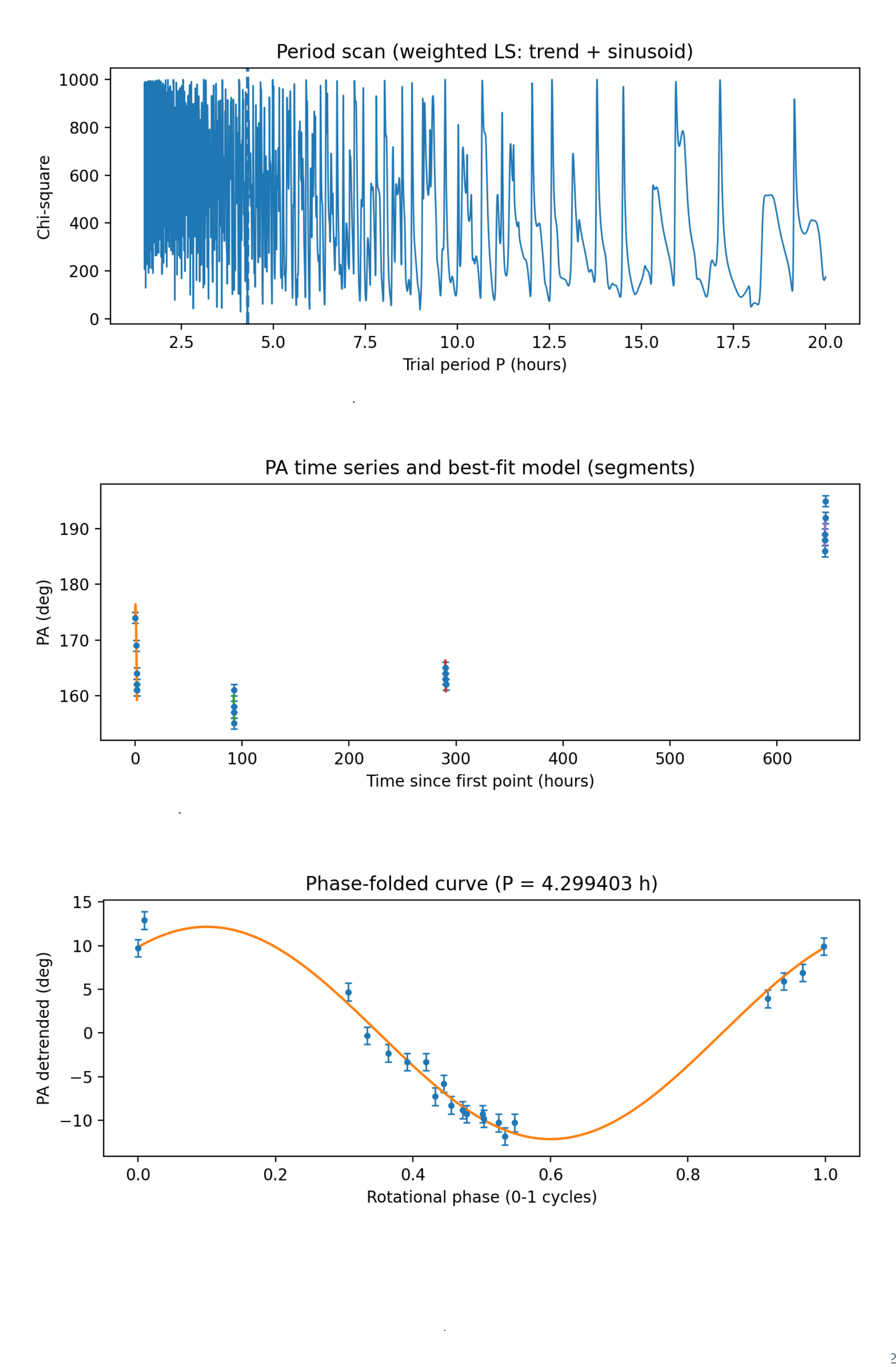}
\caption{Jet~3 diagnostic plots from the PA time series analysis: (top) period scan; (middle) PA versus time by observing segment with best-fit model segments; (bottom) detrended phase fold at $P_3=4.3$~h.}
\label{fig:jet3diag}
\end{figure*}

\paragraph{Jet 3.}
\textbf{For Jet~3, the best period is $P_3 = 4.299$~h with amplitude $A_3=12.6^\circ$ and reduced chi-square $\chi^2_\nu\simeq1.50$.
The next-best local minima (aliases) occur at $P=4.111$~h, 8.983~h, 5.983~h, 2.827~h and 3.602~h (see Figure~\ref{fig:jet3diag}).}

Given sparse sampling and evolving coma morphology, we treat $P_2$ and $P_3$ as likely aliases/harmonics or feature-specific responses rather than independent nucleus spin periods.
In a precession framework, such apparent shorter periods can arise from non-sinusoidal projection, multi-lobed emission, or morphology-dependent centroiding of the brightest streamline within a fan.

\subsection{Summary of periods}
Table~\ref{tab:periods} summarizes the periods derived from morphology and photometry.
Despite differences at the formal level, the dominant jet PA wobble and the coma photometry cluster near a characteristic post-perihelion period of $\sim7.1$~h when interpreted as an attitude/precession timescale.
The shorter Jet~1 and Jet~3 periods are included as secondary/feature-tracing signals.

\begin{deluxetable*}{llll}
\tablecaption{Summary of periodic signals used in this work.\label{tab:periods}}
\tablewidth{0pt}
\tablehead{
\colhead{Tracer} & \colhead{Data / method} & \colhead{Best period (h)} & \colhead{Notes}
}
\startdata
Dominant jet PA wobble & HST morphology; Fourier ($k\le2$) scan & $7.20\pm0.05$ & $\sigma_{\rm PA}=3^\circ$;\\
Coma photometry (small) & L92; Astrometrica $N=4$ ($r_{\rm eff}\approx6.2''$) & $7.136\pm0.001$ & $A\simeq0.311$~mag; $\sigma_{\rm jit}\simeq0.089$~mag \\
Coma photometry (large, 4 nights) & L92; Astrometrica $N=30$ ($r_{\rm eff}\approx42.2''$) & $7.142$ & RMS $\approx0.147$~mag \\
Coma photometry (large, 5 nights) & L92; Astrometrica $N=30$ ($r_{\rm eff}\approx42.2''$) & $7.158$ & RMS $\approx0.161$~mag ($k=1$) \\
Jet 1 PA & PA$(t)$ sinusoid$+$trend fit & $2.911$ & $A_1\simeq13^\circ$;\\
Jet 3 PA & PA$(t)$ sinusoid$+$trend fit & $4.299$ & $A_3\simeq12^\circ$;\\
\enddata
\end{deluxetable*}

\section{Discussion}
\subsection{Precession (NPA) interpretation of the 7.1 h period}
We interpret the $\sim7.1$~h periodicity as an attitude precession/nutation period in a non-principal-axis (NPA) rotation state, rather than as a nucleus-shaped rotational light curve.
In an NPA state, projected body axes (and any fixed active regions on the surface) can undergo a quasi-periodic wobble as the nucleus precesses about the angular-momentum vector \citep[e.g.,][]{SamarasinhaAhEarn1991,Belton2010}.
This naturally explains: (i) PA oscillations about mean directions with feature-dependent amplitudes and phases, (ii) non-sinusoidal waveforms that result from $k=2$ terms, and (iii) sensitivity to evolving collimation (fan opening) that can distort the observed waveform without requiring a change in the underlying period.

\subsection{Why coma photometry follows the precession period}
Because the measured flux is coma-dominated, the photometric modulation is best interpreted as an activity response:
precession-driven variation in the projected jet directions and collimation, changes the fraction of jet and near-nucleus dust flux within a fixed observing aperture.
A decisive test is phase locking: if the modulation is jet-driven, extrema in magnitude should correlate with the precession phase inferred from PA$(t)$, potentially with a lag representing dust transport and seeing/aperture smoothing.

\subsection{Aperture dependence as a diagnostic}
The comparison between $r_{\rm eff}\approx6.2''$ and $42.2''$ provides a direct test of a jet-driven period.
If the modulation were a bare-nucleus rotational light curve, the inferred period and waveform would be insensitive to aperture size.
Instead, the large-aperture solution exhibits degraded stationarity when the coma transitions to a fan-like morphology (December 27, 2025; see Figure~\ref{fig:jets20251227}) and shows modest shifts in preferred periods within the 6.8--7.4~h window (Table~\ref{tab:periods}).
This is expected if the measured flux is a convolution of intrinsic activity with dust transport, seeing, and aperture-dependent redistribution.

\subsection{Can multiple jets ``stabilize'' the nucleus?}
Multiple active sources can partially cancel net outgassing torques, helping to keep the angular-momentum direction relatively stable in inertial space.
However, torque cancellation does not imply an absence of rotation; rather, it can allow the nucleus to maintain a long-lived non-principal-axis state with a slowly evolving angular-momentum vector.
In this context, Jet~2---approximately anti-sunward and close to the projected rotation axis---anchors the large-scale morphology, while Jets~1 and 3 trace the precession cone through their PA oscillations.
The observed $\sim7.1$~h period is then interpreted as an attitude precession/nutation timescale that modulates both the projected jet directions and the inner-coma flux captured by fixed apertures.
\\
\\
\section{Conclusions}
We analyze data showing a coherent post-perihelion activity period in 3I/ATLAS, traced independently by inner-coma jets morphology and coma-dominated photometry.
Our main findings are:
\begin{enumerate}
\item A dominant jet-like feature in HST images shows a repeatable PA wobble with $P_{\rm jet}=7.20\pm0.05$~h from a weighted Fourier ($k\le2$) period scan, with an angular excursion of order $\sim20^\circ$.
\item Small-aperture coma photometry ($N=4$, $r_{\rm eff}\approx6.2''$) yields $P_{\rm phot}=7.136\pm0.001$~h (formal), with semi-amplitude $A\simeq0.311$~mag and additional jitter $\sigma_{\rm jit}\simeq0.089$~mag.
\item Large-aperture photometry ($N=30$, $r_{\rm eff}\approx42.2''$) gives best periods in the 7.14--7.16~h range for $k=1$ fits, while the inclusion of December 27, 2025 (fan-like jets) degrades phase stationarity and slightly shifts preferred periods.
\item We interpret the $\sim7.1$~h period as an attitude precession/nutation (NPA) signature traced by jet orientation and coma flux redistribution, rather than as a nucleus-shaped rotational light curve. These results support a non-principal-axis rotational state of 3I/ATLAS and provide new constraints on the post-perihelion activity of interstellar objects.
\textbf{To ensure robustness of our result, we also computed Lomb–Scargle periodograms (Scargle 1982) over a wide period range. The Lomb–Scargle method is particularly appropriate for irregularly sampled data and does not assume uniform cadence. The resulting periodograms show consistent peaks near 7.1 h, confirming the Fourier-based results.
Periodograms were computed over a broad range (1–20 h) without smoothing to preserve alias structures. This allows the reader to evaluate the noise characteristics and possible aliases induced by diurnal sampling. Figures 5 and 6 are presented as plots illustrating the behavior of secondary periodicities. These features are interpreted and put in relation with the primary result of this study on the dominant 7.1 h periodicity independently detected in both morphology and photometry.  We verified that the derived photometric modulation is not correlated with seeing variations. No statistically significant correlation between seeing and magnitude was found, supporting the interpretation of intrinsic nucleus-driven modulation. The agreement between independent diagnostics — jet morphology and coma photometry — strengthens the interpretation of a physical periodicity associated with nucleus dynamics.}
\end{enumerate}

AKNOWLEDGEMENTS
\textbf{A.L. was supported in part by the Harvard Black Hole Initiative (funded by GBMF and JTF) and  Galileo Project. This work was carried out using Astroart for image processing, Astrometrica for time photometry, open-source Python tools for time-series analysis and figure generation, including Astropy, Numpy, Scipy and MatPlotlib. The authors thank NASA, ESA, and STScI for the HST data used in this work.}

\startlongtable
\begin{deluxetable*}{lcccccc}
\tabletypesize{\scriptsize}
\tablewidth{0pt}
\tablecaption{Photometric measurements of 3I/ATLAS (Aperture 4 pixels)\label{tab:4pixels}}
\tablehead{
\colhead{JD} & \colhead{mag} & \colhead{Flt} & \colhead{SNR} & \colhead{ZeroPt} & \colhead{Cat} & \colhead{Design.}
}
\startdata
2461018.5459109 & 14.071 & Gr R & 37.04 & 26.859 & Gaia3 & 3I/ATLAS \\
2461018.5466227 & 13.930 & Gr R & 28.37 & 26.743 & Gaia3 & 3I/ATLAS \\
2461018.5473333 & 14.078 & Gr R & 26.12 & 26.812 & Gaia3 & 3I/ATLAS \\
2461018.5480428 & 14.316 & Gr R & 24.50 & 26.713 & Gaia3 & 3I/ATLAS \\
2461018.5487535 & 14.282 & Gr R & 31.12 & 26.902 & Gaia3 & 3I/ATLAS \\
2461018.5494641 & 13.720 & Gr R & 57.42 & 26.689 & Gaia3 & 3I/ATLAS \\
2461018.5501748 & 13.921 & Gr R & 31.55 & 26.572 & Gaia3 & 3I/ATLAS \\
2461018.5508843 & 13.833 & Gr R & 33.07 & 26.534 & Gaia3 & 3I/ATLAS \\
\enddata
\tablecomments{Observer: T. Scarmato. Obs code: L92. Telescope: 0.25m f/4.8 Newton. Exposure JD values are mid-exposure and are not corrected for light time.}
\end{deluxetable*}
\startlongtable
\begin{deluxetable*}{lcccccc}
\tabletypesize{\scriptsize}
\tablewidth{0pt}
\tablecaption{Photometric measurements of 3I/ATLAS (Aperture 30 pixels)\label{tab:30pixels}}
\tablehead{
\colhead{JD} & \colhead{mag} & \colhead{Flt} & \colhead{SNR} & \colhead{ZeroPt} & \colhead{Cat} & \colhead{Design.}
}
\startdata
2461018.5466227 & 11.489 & Gr R & 56.54 & 26.774 & Gaia3 & 3I/ATLAS \\
2461018.5473333 & 11.586 & Gr R & 67.23 & 26.823 & Gaia3 & 3I/ATLAS \\
2461018.5480428 & 11.632 & Gr R & 52.30 & 26.757 & Gaia3 & 3I/ATLAS \\
2461018.5487535 & 11.547 & Gr R & 77.70 & 26.844 & Gaia3 & 3I/ATLAS \\
2461018.5494641 & 11.632 & Gr R & 53.80 & 26.640 & Gaia3 & 3I/ATLAS \\
2461018.5501748 & 11.394 & Gr R & 63.00 & 26.833 & Gaia3 & 3I/ATLAS \\
2461018.5508843 & 11.525 & Gr R & 76.93 & 26.890 & Gaia3 & 3I/ATLAS \\
2461018.5515961 & 11.440 & Gr R & 62.64 & 26.644 & Gaia3 & 3I/ATLAS \\
\enddata
\tablecomments{Observer: T. Scarmato. Obs code: L92. Telescope: 0.25m f/4.8 Newton. Exposure JD values are mid-exposure and are not corrected for light time.}
\end{deluxetable*}

\

\startlongtable
\begin{deluxetable*}{lccccccccccc}
\tabletypesize{\scriptsize}
\tablecaption{Astrometric and photometric measurements for 3I/ATLAS obtained with the 4-pixel aperture ($\sim$6.2\arcsec).\label{astrophotometry4pixels}}
\tablewidth{0pt}
\tablehead{
\colhead{UT Date} & \colhead{UT} & \colhead{RA (deg)} & \colhead{Dec (deg)} & \colhead{$\sigma_{\rm RA}$} & \colhead{$\sigma_{\rm Dec}$} & \colhead{Mag} & \colhead{$\sigma_{\rm Mag}$} & \colhead{Seeing} & \colhead{$N_\star$}
}
\startdata
2025-12-09 & 01:06:06 & 172.752980 & +3.195692 & 0.446 & 0.396 & 14.07 & 0.292 & 5.4 & 23 \\
2025-12-09 & 01:07:08 & 172.752100 & +3.195563 & 0.516 & 0.524 & 13.93 & 0.321 & 5.8 & 12 \\
2025-12-09 & 01:08:09 & 172.751526 & +3.196003 & 0.719 & 0.609 & 14.08 & 0.296 & 5.3 & 14 \\
2025-12-09 & 01:09:10 & 172.750739 & +3.195707 & 0.604 & 0.856 & 14.32 & 0.268 & 6.3 & 18 \\
2025-12-09 & 01:10:12 & 172.749397 & +3.196257 & 0.680 & 0.864 & 14.28 & 0.314 & 5.3 & 11 \\
2025-12-09 & 01:11:13 & 172.749175 & +3.196720 & 0.283 & 0.287 & 13.72 & 0.306 & 5.6 & 22 \\
2025-12-09 & 01:12:15 & 172.748441 & +3.196915 & 0.521 & 0.445 & 13.92 & 0.267 & 3.8 & 15 \\
2025-12-09 & 01:13:16 & 172.747650 & +3.197209 & 0.617 & 0.485 & 13.83 & 0.261 & 4.0 & 14 \\
\enddata
\tablecomments{Observatory code L92; telescope 0.25~m f/4.8 Newtonian; detector CMOS ASI294MC PRO; Astrometrica v4.16.4.468; Gaia DR3 for astrometric and photometric calibration; exposure time 60 s.}
\end{deluxetable*}

\startlongtable
\begin{deluxetable*}{llcccccccc}
\tabletypesize{\scriptsize}
\tablecaption{Astrometric and photometric measurements for 3I/ATLAS obtained with the 30-pixel aperture ($\sim$42.2\arcsec).\label{astrophotometry30pixels}}
\tablewidth{0pt}
\tablehead{
\colhead{UT Date} & \colhead{UT} & \colhead{RA (deg)} & \colhead{Dec (deg)} & \colhead{$\sigma_{\rm RA}$} & \colhead{$\sigma_{\rm Dec}$} & \colhead{Mag} & \colhead{$\sigma_{\rm Mag}$} & \colhead{Seeing} & \colhead{$N_\star$}
}
\startdata
2025-12-09 & 01:07:08 & 172.752693 & +3.195580 & 1.471 & 1.511 & 11.49 & 0.309 & 3.9 & 15 \\
2025-12-09 & 01:08:09 & 172.752064 & +3.195973 & 2.039 & 1.884 & 11.59 & 0.285 & 3.7 & 11 \\
2025-12-09 & 01:09:10 & 172.751093 & +3.195436 & 1.659 & 1.644 & 11.63 & 0.271 & 4.0 & 15 \\
2025-12-09 & 01:10:12 & 172.750477 & +3.196539 & 1.424 & 1.491 & 11.55 & 0.303 & 4.0 & 18 \\
2025-12-09 & 01:11:13 & 172.749133 & +3.196693 & 0.877 & 0.880 & 11.63 & 0.301 & 4.1 & 23 \\
2025-12-09 & 01:12:15 & 172.749055 & +3.196786 & 1.761 & 1.757 & 11.39 & 0.283 & 3.8 & 22 \\
2025-12-09 & 01:13:16 & 172.747897 & +3.197240 & 1.260 & 1.229 & 11.53 & 0.258 & 3.9 & 21 \\
2025-12-09 & 01:14:17 & 172.746992 & +3.196960 & 1.343 & 1.350 & 11.44 & 0.264 & 3.8 & 29 \\
\enddata
\tablecomments{Observatory code L92; telescope 0.25~m f/4.8 Newtonian; detector CMOS ASI294MC PRO; Astrometrica v4.16.4.468; Gaia DR3 for astrometric and photometric calibration; exposure time 60 s.}
\end{deluxetable*}
\end{document}